\documentclass[aps,prl,showpacs,twocolumn,superscriptaddress,floatfix]{revtex4}
\usepackage{graphicx,amsmath,amsfonts}

\RequirePackage{xspace}

\newcommand{\BABARPubYear}     {06}
\newcommand{\BABARPubNumber}  {027}
\newcommand{\SLACPubNumber} {11817}

\input pubboard/babarsym.tex

\def\BaBar{\mbox{\slshape B\kern-0.1em{\smaller A}\kern-0.1em
    B\kern-0.1em{\smaller A\kern-0.2em R}}\xspace}
\def\babar{\BaBar}

\def\BR#1#2{\ensuremath{\cal B}(#1\to#2)\xspace}

\graphicspath{{./figures/}}

\begin{document}

\begin{flushleft}
  \babar-PUB-\BABARPubYear/\BABARPubNumber \\
  SLAC-PUB-\SLACPubNumber \\
\end{flushleft}

\smallskip

\title{ \large \bf \boldmath
Observation of \FourS decays to $\pipi\OneS$ and
$\pipi\TwoS$}

\smallskip

%
\author{B.~Aubert}
\author{R.~Barate}
\author{M.~Bona}
\author{D.~Boutigny}
\author{F.~Couderc}
\author{Y.~Karyotakis}
\author{J.~P.~Lees}
\author{V.~Poireau}
\author{V.~Tisserand}
\author{A.~Zghiche}
\affiliation{Laboratoire de Physique des Particules, F-74941 Annecy-le-Vieux, France }
% ...
\collaboration{The \babar\ Collaboration}
\noaffiliation

\begin{abstract}
We present the first measurement of
\FourS\ decays to \pipi\OneS\ and \pipi\TwoS
based on a sample of 230$\times10^6$ \FourS\ mesons collected with
the \babar\ detector. 
We measure the product branching fractions
${\cal B}(\FourS\to \pipi\OneS)\times{\cal B}(\OneS\to\mumu)=( 2.23\pm0.25_{stat} \pm0.27_{sys} )\times 10^{-6}$ and
${\cal B}(\FourS\to \pipi\TwoS)\times{\cal B}(\TwoS\to\mumu)=(1.69\pm0.26_{stat}\pm0.20_{sys})\times 10^{-6}$,
from which we derive the partial widths 
${\Gamma}(\FourS\to \pipi\OneS)=(1.8\pm0.4)$~keV and
${\Gamma}(\FourS\to \pipi\TwoS)=(2.7\pm0.8)$~keV.

\end{abstract}


\maketitle

The \FourS\ meson is known to decay predominantly to \BB, with small, but as yet 
unobserved, decays to other bottomonium states or to light hadrons.
Partial widths for hadronic transitions in heavy quarkonia have been extensively studied
both experimentally and theoretically over the past decades~\cite{Brambilla:2004wf}.
In particular, the values of the partial widths for  dipion transitions 
between vector states $\psi(2S)\to\pipi J/\psi$ and $\Upsilon(mS)\to\pipi\Upsilon(nS)$, 
where the principal quantum number $m>n$,
can be related to the radial wave function within the framework of the QCD multipole 
expansion~\cite{Kuang:1981se}.
This picture may be significantly altered by mixing and coupled channel 
effects~\cite{Boh} when states are close to the threshold for open 
charm or bottom production. Hence these states are the ideal laboratory
to investigate these effects.  
Exclusive non-$D\Dbar\xspace$ decays of the $\psi(3770)$ (believed to be predominantly $^3D_1$)
have recently been observed~\cite{Bai:2003hv,Adam:2005mr,Coan:2005ps}, but only 
upper limits have been published for exclusive non-\BB\ decays of the
\FourS~\cite{Glenn:1998bd}.

We search for the decays $\Upsilon(4S)\to\pipi\Upsilon(nS)$, where $n=1,\,2$~\cite{FN}, using a sample of
$230\times10^6$ \FourS 
events corresponding to an integrated luminosity of 211$\,\invfb$ acquired near the peak of the \FourS resonance 
with the PEP-II asymmetric-energy $\epem$ storage rings at SLAC. 
An additional 22$\,\invfb$ sample collected approximately 40$\,\mev$ below the resonance 
is used as a control sample.

The \BaBar\ detector is described in detail elsewhere~\cite{babar-detector}; here we summarize only
the features relevant to this analysis:
charged-particle momenta are measured in a tracking system consisting 
of a five-layer double-sided silicon vertex tracker (SVT) and a 
40-layer central drift chamber (DCH), both situated in a 1.5-T axial 
magnetic field. Charged-particle identification is based on
the \dedx measured in the SVT and DCH,
and on a measurement of
the photons produced in the synthetic fused-silica bars of the
ring-imaging Cherenkov detector (DIRC).
 A CsI(Tl) electromagnetic 
calorimeter (EMC) is used to detect and identify photons and 
electrons, while muons are identified in the instrumented 
flux return of the magnet (IFR).

An $\Upsilon(mS)\to\pipi\Upsilon(nS)$ transition, denoted by $mS\to nS$, 
is detected by reconstructing the $\Upsilon(nS)$ meson
via its leptonic decay to $\mumu$.  The sensitivity to $4S\to nS$ transitions is much smaller
in the $\pipi\epem$ final state due to the presence of larger backgrounds, and to 
a trigger-level inefficiency introduced by the prescaling of Bhabha scattering events.
Data collected at a nominal center-of-mass energy $\sqrt{s}$ near 10.58~\gev include 
$3S\rightarrow nS$ ($n=1,\,2$) and $2S \rightarrow 1S$ events from initial state radiation (ISR) production 
that are used as control samples.
The signature for $mS\to nS$ transition events, where the $nS$ decays to muons,
is a $\mumu$ invariant mass, $M_{\mu\mu}$, that is compatible with the known mass~\cite{PDG} of the $\Upsilon(nS)$ 
resonance, $M(nS)$, and an invariant mass difference $\Delta M=M_{\pi\pi\mu\mu}-M_{\mu\mu}$ 
that is compatible with $M(mS)-M(nS)$. The r.m.s. values of the reconstructed $\Delta M$ and $M_{\mu\mu}$ distributions are, respectively, 
$\approx 7\,\mevcc$ and $\approx 75\,\mevcc$.
The center-of-mass momentum $p^*_{cand}$ should be compatible with
0 for $4S\to nS$ candidates, or with $\left(s-M^2(mS)\right)/(2\sqrt{s})$ for $mS\to nS$ candidates from ISR.

Simulated Monte Carlo (MC) events are generated 
using the EvtGen package ~\cite{EVTGEN}.
The angular distribution of generated dilepton decays incorporates 
the $\Upsilon(nS)$ polarization, 
while dipion transitions are generated according to phase space. 
These events are passed through a detector simulation based on GEANT4~\cite{GEANT},
and analyzed in the same manner as data.
The events in the data sample whose values of 
$\Delta M$ and $M_{\mu\mu}$ are
within 60~\mevcc and 300~\mevcc, respectively, of the values
expected for any known $mS\to nS$ transition were not examined until the
event selection criteria were finalized.  Events outside these regions
were used to understand the background.

We select events having at least 4 charged tracks with a polar angle $\theta$ within
the fiducial volume of the tracking system (0.41$<\theta<$2.54~rad).
Each muon candidate is required to have a center-of-mass momentum  greater than 
4\gevc, and to be compatible with the muon hypothesis based on the energy deposited in the EMC 
and the hit pattern in the IFR along the track trajectory.
A dipion candidate is formed from a pair of oppositely charged tracks.
The two pion candidates are each required to have a transverse momentum greater than $100\,\mevc$. 
The dimuon and the dipion are constrained to a common vertex, and the vertex fit is required to have a
$\chi^2$ probability larger than $10^{-3}$.

\begin{figure}[tb]
 \begin{center}
   \includegraphics[width=8.8cm]{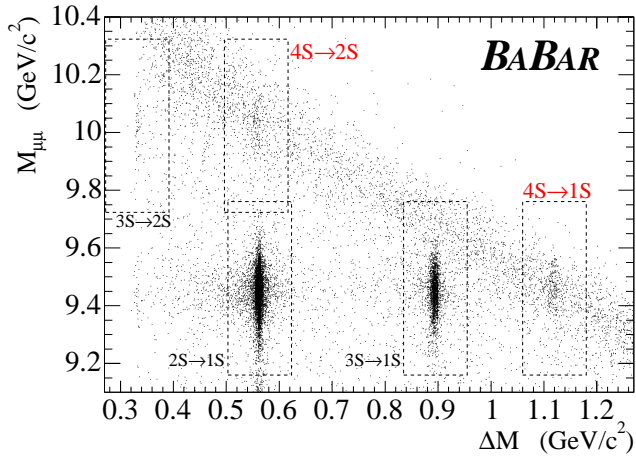}
\end{center}
\caption{
The $M_{\mu\mu}$ vs. $\Delta M$  
distribution. 
Dashed lines delimit the regions where $\Delta M$ and $M_{\mu\mu}$ are within $\pm 60\,\mevcc$ and $\pm 300\,\mevcc$, respectively, of the values expected
for an $mS\to nS$ transition. The remaining region is used to model background. 
The text discusses the features seen in the data.}
\label{fig:DeltaMall}
\end{figure}

A large fraction of the background is due to $\mumu\gamma$ events where a photon converts
in the detector material.
To reduce this background we apply an ``electron veto'',
rejecting events where any of the following is true: either of the two pion 
candidates is positively identified as an electron;
the $e^+e^-$ invariant mass of the two 
charged tracks associated with the pion candidates satisfies $M_{ee}<100\,\mevcc$;
or the dipion opening angle satisfies $\cos{\theta_{\pipi}}>0.95$.
The distribution of $\Delta M$ vs $M_{\mu\mu}$ for the final sample is shown in Fig.~\ref{fig:DeltaMall}.
The clusters of events in the boxes centered at
$\left (\Delta M,M_{\mu\mu}\right)=(1.120,9.460)\,\gevcc$ and $(0.558,10.023)\,\gevcc$ constitute, respectively, 
the first observation of $4S\to 1S$ and of 
$4S\to 2S$ transitions. We also observe signals for $2S\to 1S$, $3S\to 2S$, and $3S\to 1S$ from ISR at 
$\left (\Delta M,M_{\mu\mu}\right)=(0.563,9.460)\,\gevcc$, $(0.332,10.023)\,\gevcc$, and $(0.895,9.460)\,\gevcc$ respectively. 
The diagonal band is predominantly due to $\mu\mu\gamma$ events, while the cluster at $\left (\Delta M,M_{\mu\mu}\right)=(0.332,9.460)\,\gevcc$
is due to 
$\Upsilon(3S)\to\pipi\Upsilon(2S)$ decays, where $\Upsilon(2S)\to \Upsilon(1S)\,X$.

\begin{figure}[tb]
\begin{center}
\begin{tabular}{cc}  
 \includegraphics[width=4.3cm]{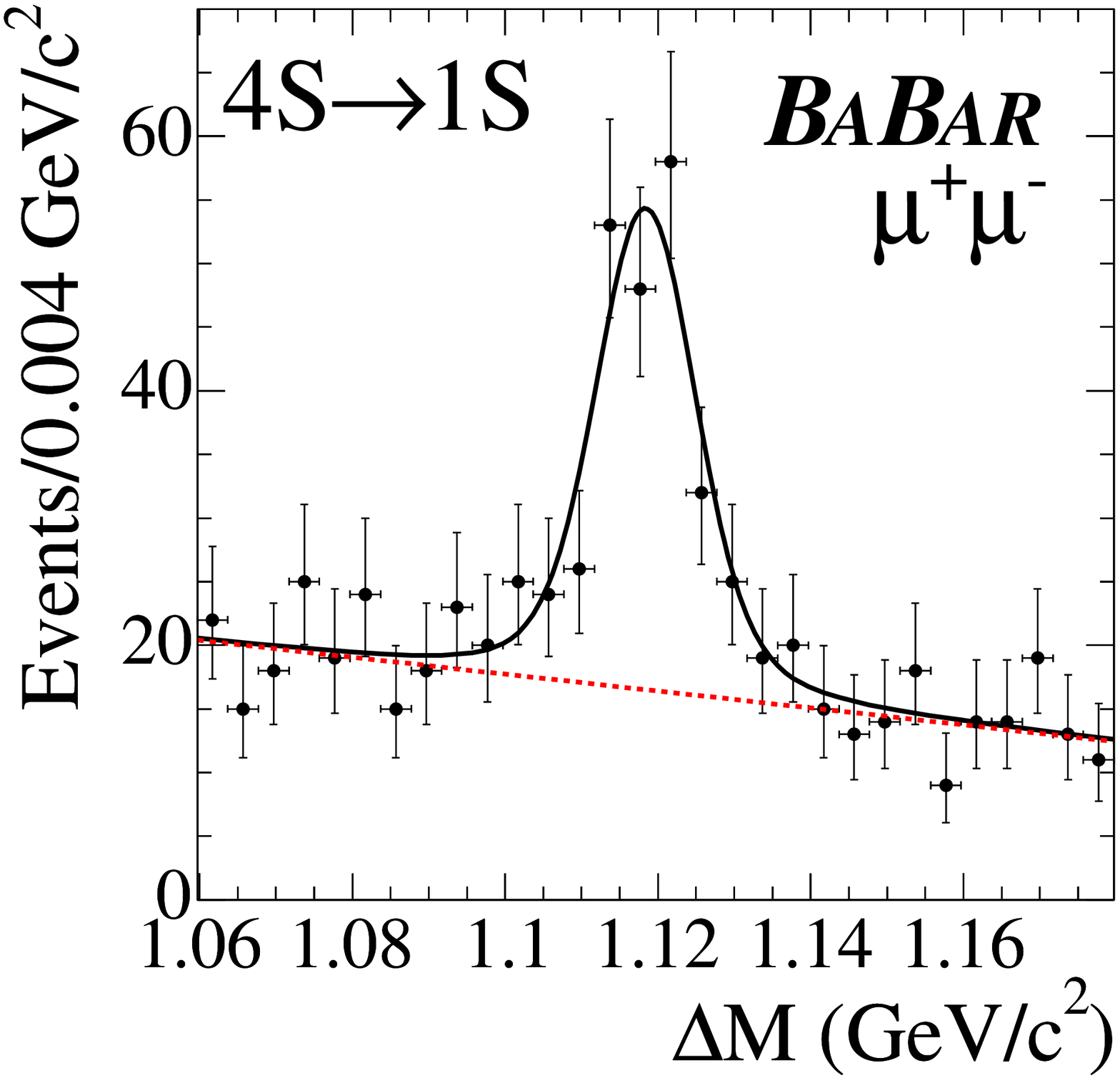} &
 \includegraphics[width=4.3cm]{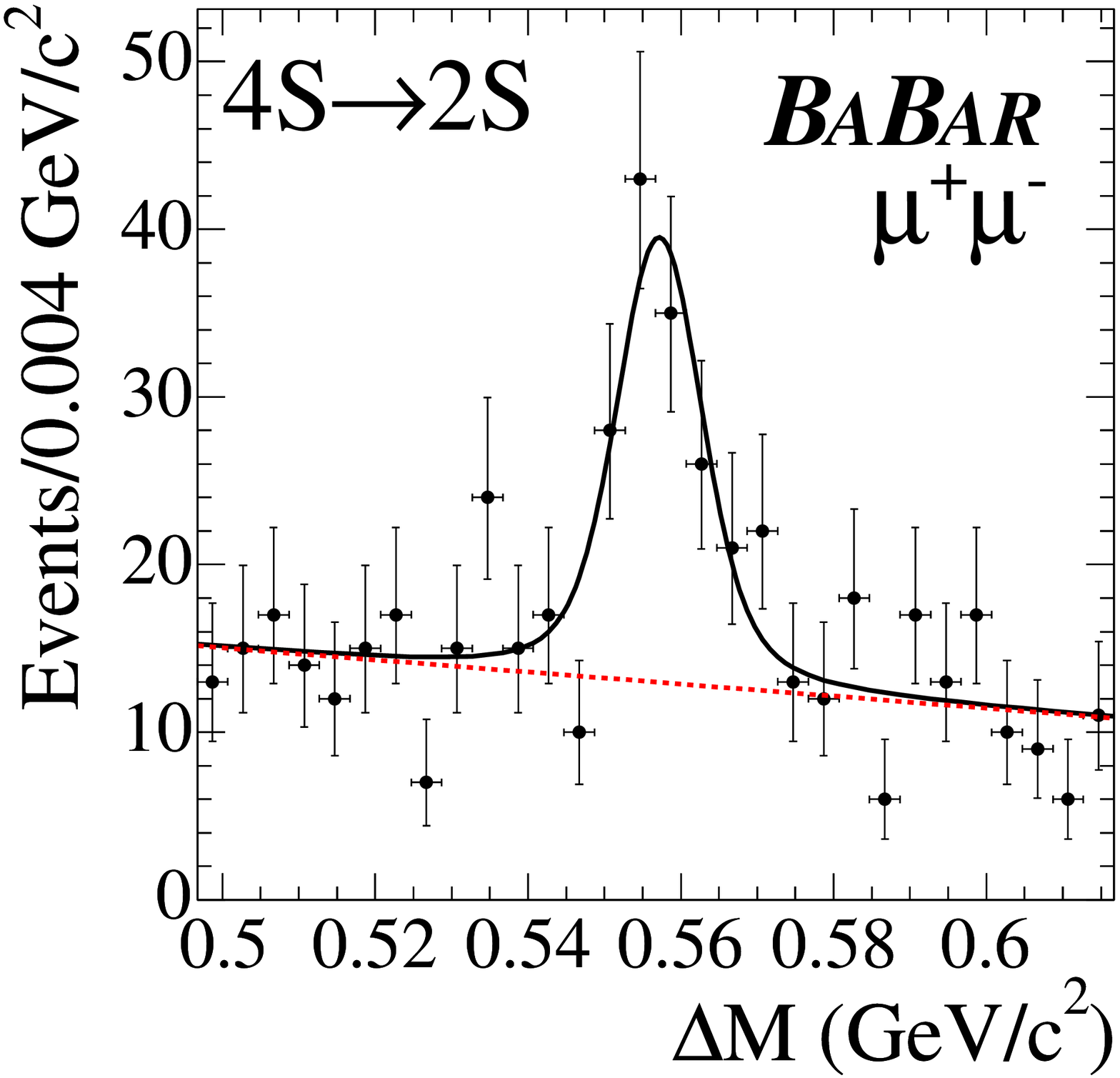}\\
 \includegraphics[width=4.3cm]{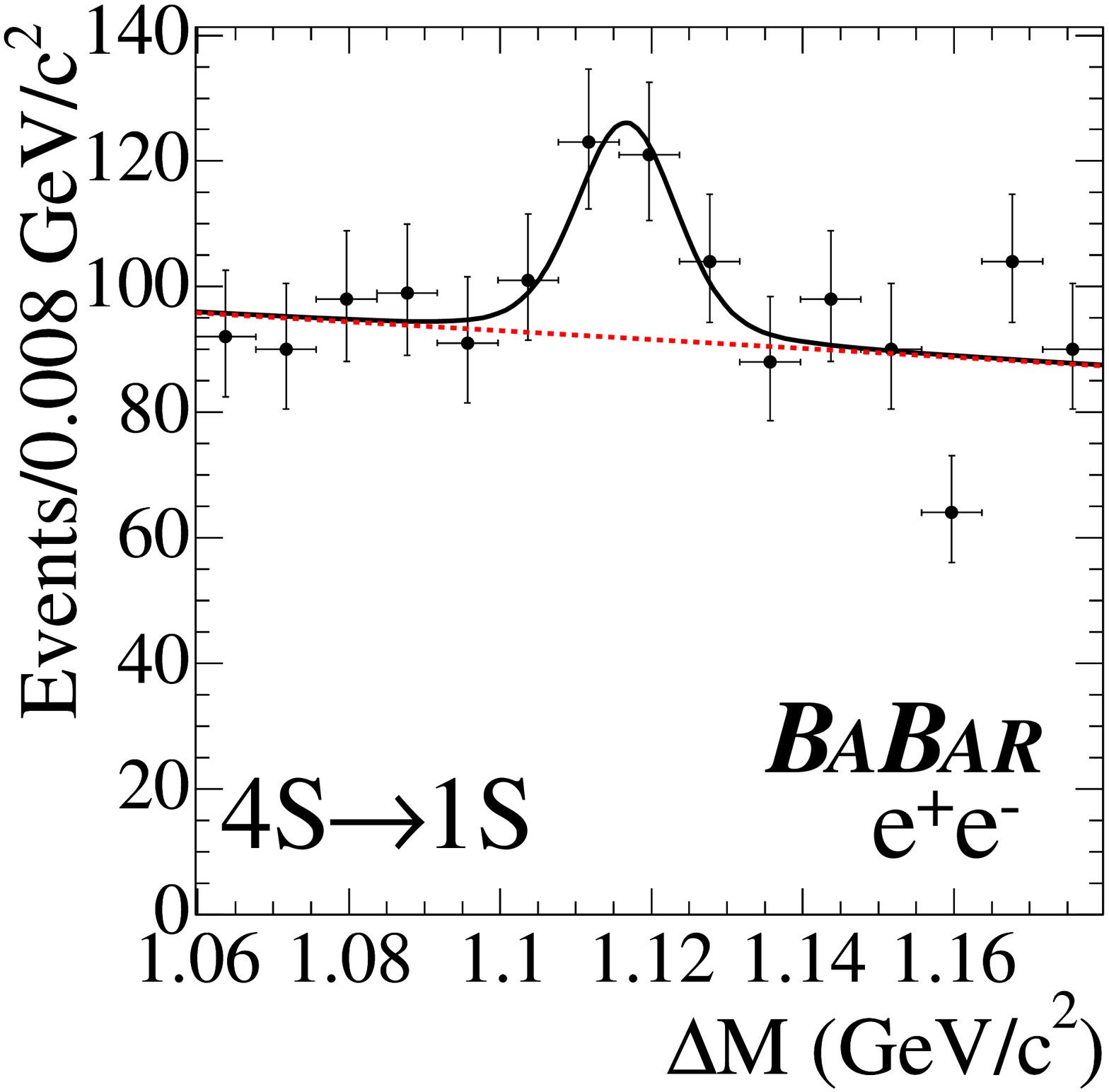} &
 \includegraphics[width=4.3cm]{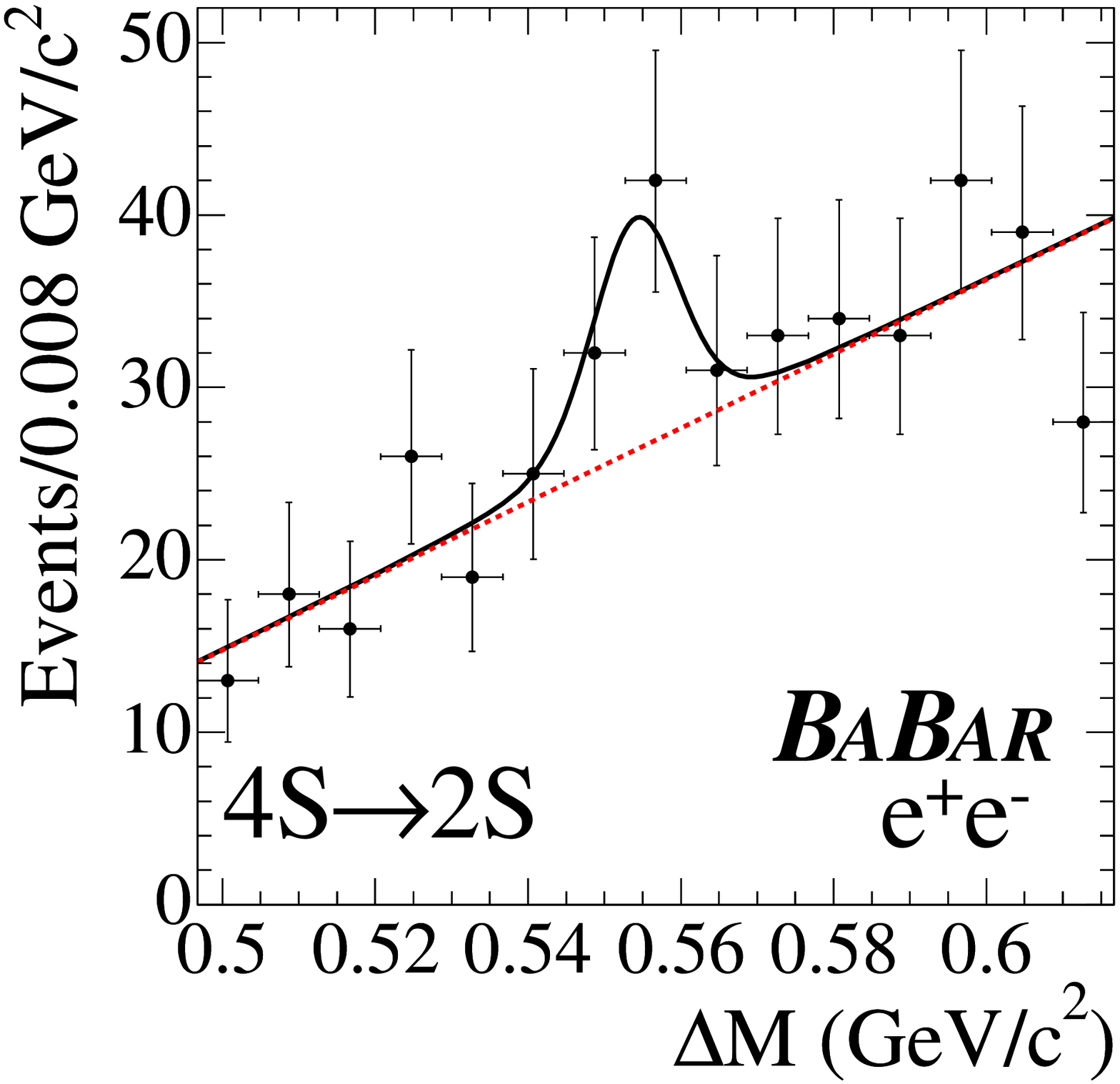}\\
\end{tabular}
\end{center}
\caption{
The $\Delta M$ distribution for events with $\vert M_{\ellell}-M(1S)\vert <200 \mevcc$ (left)
and  $\vert M_{\ellell}-M(2S)\vert <150 \mevcc$ (right).
The solid lines show the best fit to the data.  Dashed lines show the background contribution.
The two upper plots are for $\pipi\mumu$ candidates and the two lower plots for $\pipi\epem$ candidates.}
\label{fig:Fits}
\end{figure}

The number of signal events $N_{sig}$ is extracted by an unbinned extended maximum likelihood fit to the $\Delta M$ distribution for
events with $p^*_{cand}<200\,\mevc$ and $\vert M_{\mu\mu}-M(1S)\vert < 200\,\mevcc$ for the $4S\to 1S$ mode or $\vert M_{\mu\mu}-M(2S)\vert < 150\,\mevcc$ for the
$4S\to 2S$ mode 
(Fig.~\ref{fig:Fits}). In each case, the background is parametrized as a linear function, and the signal 
as the convolution of a Gaussian with standard deviation $\sigma$ and a Cauchy
function with width $\Gamma$, which is found to adequately describe the non-Gaussian tails of the $\Delta M$ distribution. The values for $\sigma$ and $\Gamma$ are, for each
mode, fixed to the values determined from a fit to a MC signal sample subjected to the detector simulation and reconstruction algorithms. 
We verify  that the experimental $\Delta M$ resolution 
is well described by the MC
for $2S\to 1S$ and $3S\to nS$ ($n=1,\,2$) ISR samples.
The values of $\Delta M$ returned by the fit,  $1.1185\pm0.0009\,\gevcc$ and $0.5571\pm0.0010\,\gevcc$, where the errors are statistical only,
are in excellent agreement with the world averages $M(4S)-M(1S)=1.1197\pm0.0035\,\gevcc$ and 
$M(4S)-M(2S)=0.5567\pm0.0035\,\gevcc$~\cite{PDG}.  These values cannot be interpeted as a new measurement of the \FourS\ mass,
since data come from the peak of the resonance and not from a scan of it. 
The cuts described above are also applied to 
$\pipi\epem$ candidates, with the additional
requirement on the polar angle of the electron, $\theta(e^-)>0.75\,$radians, to reject Bhabha events.  
The fits to the electron samples
are also shown in Fig.~\ref{fig:Fits}, and give yields and $\Delta M$ values consistent with expectations
based on the fits to the muon samples. 
 
The significance, estimated from the likelihood ratio $n\sigma\simeq\sqrt{2\log\left[{{\cal L}(N_{sig})/{\cal L}(0)}\right]}$
between a fit that includes a signal
function and a fit with only a background hypothesis, is 10.0$\sigma$ for the $4S\to 1S$
and 7.3$\sigma$ for the $4S\to 2S$ in the $\pipi\mumu$ final states. The significance
of the signals in the $\pipi\epem$ final states is 3.6$\sigma$ and 2.5$\sigma$ for $4S\to 1S$ and $4S\to 2S$, 
respectively.

The event selection efficiency $\epsilon_{sel}$ is determined using the MC samples.  The largest source of systematic
uncertainty (10\%) 
is due to the unknown distribution of the dipion invariant mass in the $\FourS\to\pipi\Upsilon(nS)$ transition, and is
estimated by comparing the acceptance for a phase space distribution to that obtained using
the QCD multipole model~\cite{Kuang:1981se}. The second largest source of systematic uncertainty is due to uncertainty in the track reconstruction
 efficiency, which is 1.3\% per track, resulting in a 5.2\%  uncertainty in $\epsilon_{sel}$.
The systematic uncertainties associated with the event selection (4.3\%)
and muon identification (1.4\%) criteria  are estimated by comparing the efficiency of each selection criterion 
determined from MC samples to the corresponding efficiency
measured with the ISR control samples. 
We have also considered the systematic uncertainties due to the choice of signal and background parametrizations 
by using different functions or different parameters, and the systematic uncertainties due to the choice of the
fit range. The contributions from these sources are negligible in comparison to the previously mentioned sources.

\begin{table}[t!]
\begin{center}
\caption{\label{tab:Results} Number of signal events, significance, efficiency and measured values of the products of branching ratios for
the $4S\to nS$ transitions. The error on the efficiency is obtained adding in quadrature the
systematic uncertainties. The errors on the product branching fractions are statistical and systematic respectively} 
\begin{ruledtabular}
\begin{tabular}{lcccc}
Transition& $N_{sig}$& significance & $\varepsilon_{sel}$  & ${\cal B}_{4S\to nS}\times{\cal B}_{nS\to \mu\mu}$ \\
          &            &              & (\%) & ($10^{-6}$)  \\
\hline
$4S\to 1S$ & 167$\pm$19 & 10.0$\sigma$ & 32.5$\pm$3.9 & 2.23$\pm$0.25$\pm$0.27 \\
$4S\to 2S$ & 97$\pm$15  &  7.3$\sigma$ & 24.9$\pm$3.0 & 1.69$\pm$0.26$\pm$0.20 \\
\end{tabular}
\end{ruledtabular}
\end{center}\end{table}

The product branching fraction (Table~\ref{tab:Results}) is
determined from the $\pipi\mumu$ sample using:
\begin{eqnarray}
{\cal B}\left(\Upsilon(4S)\to \pipi\Upsilon(nS)\right)\times&&\nonumber \\
 {\cal B}\left(\Upsilon(nS)\to \mumu\right)& = &
\frac{N_{sig}}{\varepsilon_{sel}\,N(4S)},\label{eq:BR}
\end{eqnarray}
where $N(4S)=(230.0\pm2.5)\times10^6$ is the total number of \FourS\ mesons produced.

The event yields observed for $3S\to nS$ and $2S\to 1S$
are compatible with PDG-averaged values of the ISR cross section and branching fractions 
for those resonances.  
The number of signal events 
observed in the $\pipi\epem$ final state is
compatible with the branching fractions
we measure in the $\pipi\mumu$ sample. No $4S\to nS$ signal is observed for 
$\pipi\mumu$ or $\pipi\epem$ final states
in the data collected at center of mass energies 40~\mev below the $\FourS$ resonance.

\begin{figure}[tb]
 \begin{center}
\begin{tabular}{cc}  
 \includegraphics[width=4.35cm]{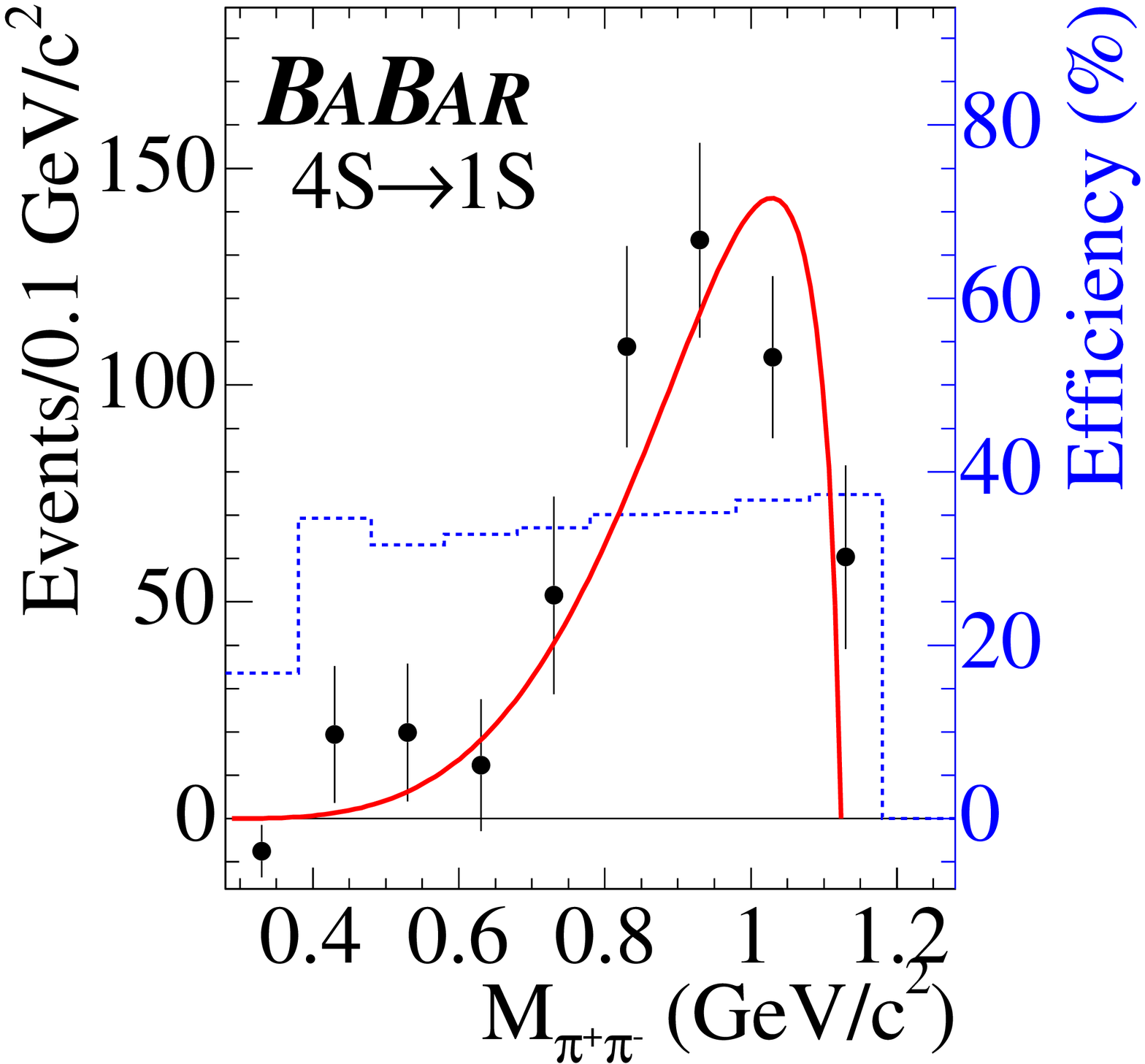} &
 \includegraphics[width=4.35cm]{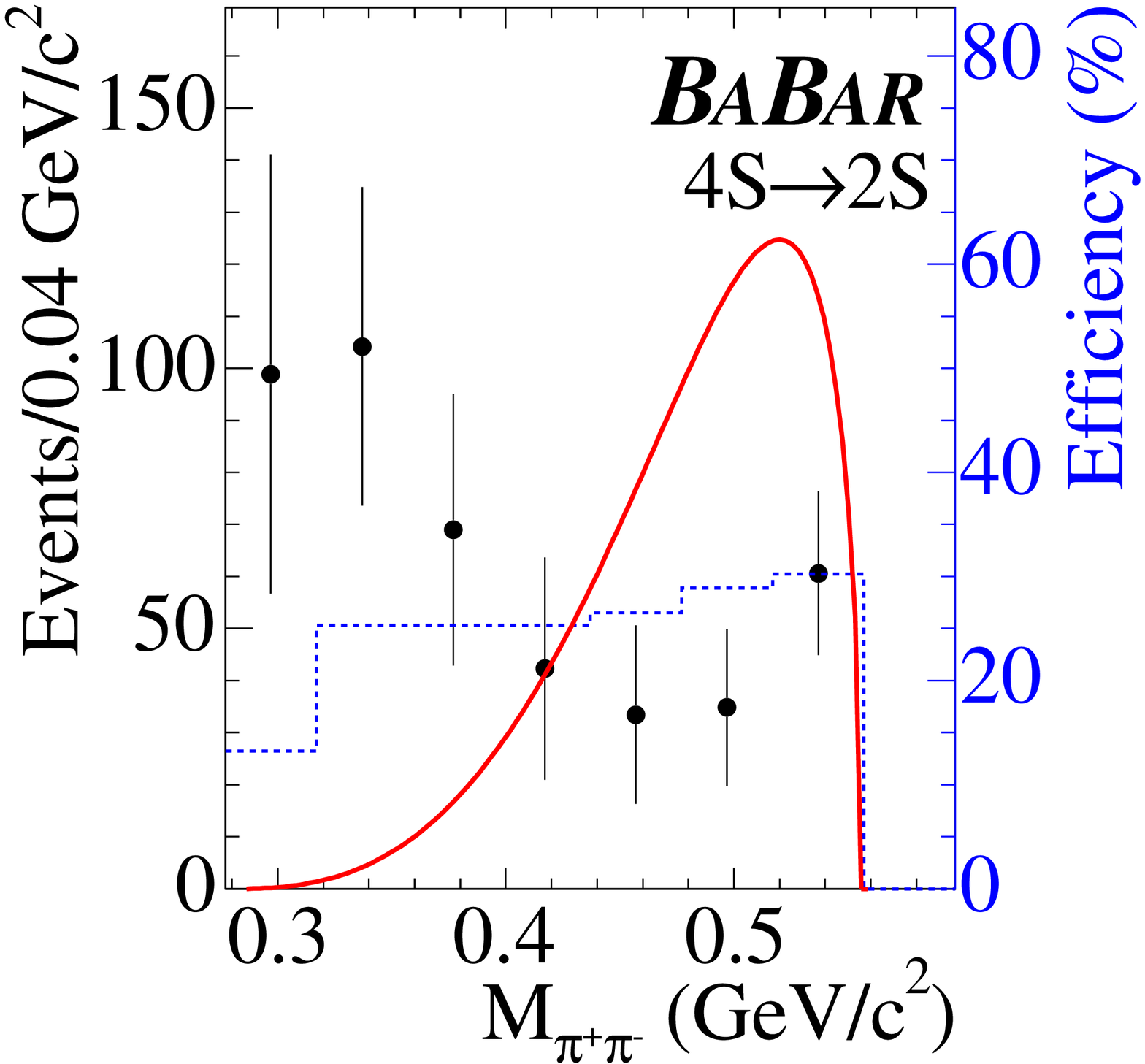}
\end{tabular}
\end{center}
\caption{The efficiency-corrected $M_{\pipi}$ distribution for $4S\to 1S$ transition (left)
and $4S\to 2S$ transition (right). The solid line shows the distribution predicted in Ref.~\cite{Kuang:1981se}. The dotted histogram shows the selection efficiency
in each bin. The experimental resolution in $M_{\pipi}$ is less than $ 5\,\mevcc$, much smaller
than the bin size.}
\label{fig:Mpipi}
\end{figure}

The dipion invariant mass distribution, $M_{\pipi}$ (Fig.~\ref{fig:Mpipi}), is determined by fitting
the $\Delta M$ distribution in equal intervals of $M_{\pipi}$, and dividing the number of signal events in each interval by the corresponding selection efficiency.
The  measured distribution for the $4S\to1S$ transition has a shape similar to
the prediction of the Kuang-Yan model~\cite{Kuang:1981se}. 
This model provides a good description of the observed distributions for $2S\to 1S$, $3S\to 2S$, and also
$\psi(2S)\to\pipi J/\psi$, but fails to describe the $3S\to 1S$ distribution. 
Our measured distribution for the $4S\to 2S$ transition has a marked enhancement at low $M_{\pipi}$ that is 
incompatible with this model.

The $4S\to nS$ branching ratios and partial widths can be derived
using the world
average values for ${\cal B}\left(\Upsilon(nS)\to \mumu\right)$  \cite{PDG} and a recent \babar\ measurement of $\Gamma(\FourS)$~\cite{Aubert:2004pw}.
We obtain
\begin{eqnarray}
&{\cal B}\left(\Upsilon(4S)\to \pipi\Upsilon(1S)\right)=(0.90\pm0.15)\times10^{-4},&\nonumber\\
&{\cal B}\left(\Upsilon(4S)\to \pipi\Upsilon(2S)\right)=(1.29\pm0.32)\times10^{-4},&\nonumber\\
&{\Gamma}(\FourS\to \pipi\OneS)=(1.8\pm0.4)~{\rm keV},&\nonumber
\end{eqnarray}
and
$${\Gamma}(\FourS\to \pipi\TwoS)=(2.7\pm0.8)~{\rm keV}.$$ 
We add in quadrature the statistical and systematic uncertainties on the derived quantities.
With the most recent CLEO measurement of
${\cal B}\left(\Upsilon(2S)\to \mumu\right)$~\cite{Adams:2004xa}, we obtain 
smaller values: ${\cal B}\left(\Upsilon(4S)\to \pipi\Upsilon(2S)\right)=(0.83\pm0.16)\times10^{-4}$ and
  ${\Gamma}(\FourS\to \pipi\TwoS)=(1.7\pm0.5)$~keV.

The branching fractions are compatible with previous upper 
limits on these decays~\cite{Glenn:1998bd}.
The $\FourS$ partial widths are within the range spanned by 
other dipion transitions
in the $b\bar b$ system~\cite{PDG}: $\Gamma(\Upsilon(2S)\to \pipi\Upsilon(1S))=(8.1\pm2.1)\,$keV; 
$\Gamma(\Upsilon(3S)\to \pipi\Upsilon(1S))=(1.2\pm0.2)\,$keV; 
$\Gamma(\Upsilon(3S)\to \pipi\Upsilon(2S))=(0.6\pm0.2)\,$keV.

In conclusion, we 
 measure 
\begin{eqnarray}
{\cal B}\left(\Upsilon(4S)\to \pipi\Upsilon(1S)\right)\times{\cal B}\left(\Upsilon(1S)\to \mumu\right) =&& \nonumber\\
  (2.23\pm0.25\pm0.27)\times10^{-6}&&\nonumber
\end{eqnarray}
and 
\begin{eqnarray}
{\cal B}\left(\Upsilon(4S)\to \pipi\Upsilon(2S)\right)\times{\cal B}\left(\Upsilon(2S)\to \mumu\right) =&&\nonumber\\
 (1.69\pm0.26\pm0.20)\times10^{-6}&&.\nonumber
\end{eqnarray}

The dipion invariant mass distribution is measured for $\FourS\to\pipi\OneS$ and
$\FourS\to\pipi\TwoS$ transitions; the latter is found to be incompatible with 
predictions from QCD multipole expansions.

We are grateful for the excellent luminosity and machine conditions
provided by our \pep2\ colleagues, 
and for the substantial dedicated effort from
the computing organizations that support \babar.
The collaborating institutions wish to thank 
SLAC for its support and kind hospitality. 
This work is supported by
DOE
and NSF (USA),
NSERC (Canada),
IHEP (China),
CEA and
CNRS-IN2P3
(France),
BMBF and DFG
(Germany),
INFN (Italy),
FOM (The Netherlands),
NFR (Norway),
MIST (Russia), and
PPARC (United Kingdom). 
Individuals have received support from CONACyT (Mexico), 
Marie Curie EIF (European Union),
the A.~P.~Sloan Foundation, 
the Research Corporation,
and the Alexander von Humboldt Foundation.


\begin{thebibliography}{99}
\bibitem{FN}The $\Upsilon(4S)\to\pi\pi\Upsilon(3S)$ transition is kinematically not allowed.
\end{thebibliography}
\end{document}